\documentclass{jetpl}
\twocolumn
\usepackage{amsmath,bm,epsfig}
\def\REF#1{Eq.~\Ref{#1}}
     \def\Ref#1{(\ref{#1})}  \def\<{\left\langle}
     \def\>{\right\rangle}

\title{On the problem of catastrophic relaxation   in superfluid $^3$He-B }

\rtitle{Catastrophic relaxation of homogeneous precession in
$^3$He-B}

\sodtitle{On the problem of catastrophic relaxation  in superfluid
$^3$He-B}

\author{Yu.\,M.\,Bunkov$^{a}$
\/\thanks{e-mails: ~~~~~~~~ ~~~~~~~~~~ ~~~~~~~~~~
~~yuriy.bunkov@grenoble.cnrs.fr, ~~~~~~~~ ~~~~~~~~~~
victor.lvov@weizmann.ac.il,    ~~~~~~~~~~ volovik@boojum.hut.fi},
V.\,S.\,L'vov$^{b,~c}$, G.\,E.\,Volovik$^{c,~d }$}

\rauthor{Yu.\,M.\,Bunkov, V.\,S.\,L'vov, G.\,E.\,Volovik}

\sodauthor{Bunkov, Lvov, Volovik}

\address{$^{a}$ Centre de Recherches sur les Tr\`es Basses Temp\'eratures,
   CNRS, BP166, 38042, Grenoble, France
\\~\\$^{b}$ Department of
Chemical Physics, The Weizmann Institute of Science, Rehovot
76100, Israel
\\~\\
$^{c}$ Low Temperature Laboratory, Helsinki University of
Technology,
P.O.Box 2200, FIN-02015, HUT, Finland\\~\\
$^{d}$ Landau Institute for Theoretical Physics RAS, Kosygina 2,
117334 Moscow, Russia}

\dates{15 August  2006}

\abstract{ In this  Letter we discussed  the parametric
instability of texture of homogeneous (in time) spin precession,
explaining how spatial inhomogeneity of the  texture may change
the threshold of the instability in  comparison with   idealized
spatial homogeneous case, considered in our~JETP Letter
\textbf{83}, 530 (2006).  This discussion is inspired   by
critical comment of I.A. Fomin (cond-mat/0606760) related to
the above questions. In addition we considered here results of
direct numerical simulations of the full Leggett-Takagi equation
of motion for magnetization in $^3$He-B and experimental data for magnetic
field dependence of the catastrophic relaxation, that provide
solid support of the theory of this phenomenon, presented in our
2006 JETP Letter.}


\begin{document}

\maketitle

\noindent
   {\bf Introduction}. This Letter is inspired by critical
   comment~\cite{Fcr} on our JETP Letter ``Solution of the problem of
   catastrophic relaxation of  homogeneous spin precession in
   superfluid $^3$He-B"~\cite{BLV}.
The self sustained and long-lived spin precession with the
coherent phase across the whole precessing domain -- the
Homogeneously Precessing Domain (HPD) -- is a unique feature of
superfluid $^3$He-B. This phenomenon bears all the ingredients of
spin superfluidity \cite{HPD}.  Later it was found that at low
temperatures, below $\sim (0.4-0.5) T_c$, the spin superfluidity
experiences abrupt instability, called the catastrophic relaxation
\cite{CatHPD}.

After many attempts it was finally recognized that the origin of
the catastrophe is the parametric (Suhl) instability
\cite{SF,BLV}.  Two competing contributions to the increment of
the parametric instability were suggested:
\begin{equation}
\label{sum}
V(\bm r)=V_{_{\rm SF}}(\bm r)+V_{_{\rm BLV}}(\bm r)\ .
\end{equation}
The ``intrinsic'' contribution $V_{_{\rm SF}}$, suggested in the
Sourovtsev-Fomin (SF) Ref.~\cite{SF},  is due to the anisotropy of
spin-wave velocity and it comes from the region where the
configuration of spin and orbital vectors   $\bm S$ and $\bm L$ is
canonical, the so-called the Brinkman-Smith (BS) mode of
precession.  Suggested in our Bunkov-L'vov-Volovik (BLV)
Ref.~\cite{BLV} contribution $V_{_{\rm BLV}}$ is due the
oscillating spin-orbit energy and it only comes from the region
where precession deviates from the BS mode.

Comment~\cite{Fcr} clarifies aspects of the problem that may cause
misunderstanding and thus require more detailed explanation and
even further development. This is the subject of our Letter, which
is organized as follows. Sections 1-3 are devoted to questions,
arose in~\cite{Fcr}, which are mainly related to the problem of
parametric instability  in the case of the spatially inhomogeneous
precession. In Sec.~4 we consider extension of our theory to other
precessing states. In Secs.~5 and 6 we discuss results of direct
numerical simulations of the full Leggett-Takagi equations for
magnetization and of laboratory study of the magnetic field
dependence of the catastrophic relaxation, that provide additional
support of our theory of this phenomena, presented in
Ref.~\cite{BLV}, and demonstrates that the SF mechanism of the
instability~\cite{SF} is in contradiction
  with the experiments.

\noindent {\bf 1. Stability of the basic reference  state}. This
is the first question of Com.~\cite{Fcr} that  is useful to
clarify.  As stated in~\cite{Fcr}, the spin-orbit potential energy
as a function of $S_z$ has concave or  the convex form shape,
depending on orientations of $\bm L$ and $\bm S$ and therefore  in
some cases  the spatially homogeneous precession can be unstable
\cite{F2}. For example, the conventional homogeneous precession is
unstable in $^3$He-A \cite{HeA}. However, this is not applicable
to a ``texture" with some profile of the orientations of the
orbital and spin moments, $\bm L$ and $\bm S$, that realizes the
minimum of the sum of the gradient and the spin-orbit energies
(see Ref.~\cite{text} for the texture in slab geometry and
Ref.~\cite{Wetting} for the texture under discussion). The texture
is stable  with respect to perturbations with the same rotation
frequency  $\omega_L$. This
   texture, shown in Fig.~1 of our Ref~\cite{BLV},  serves as the
   basic  reference state, stability of which with respect to
   parametric excitation of the spin waves with frequency $
\omega_s(\pm \bm k)= \omega_L/2\,,$
   was studied in our
   Ref.~\cite{BLV}.   Notice again, that \emph{the texture is stable
   with respect to perturbations with the frequency} $\omega_L$.
   This is the answer to the  first question of~\cite{Fcr}: ``Why
   the same mechanism  which is responsible for the instability of the
uniform precession in
$^3$He-A
    can be disregarded?".\\
    \noindent
   {\bf 2. Parametric instability  in spatially inhomogeneous
   media}.
The second statement of   Com.~\cite{Fcr} is trivial: because  the
spatial variations of the parametric instability increment $V$
``is not small\  \dots\ ,    two ways of averaging [ $\exp (\< V\>
t)$ and $\< \exp ( V t)\> $ ] can give very different results".
The relevant question  here: ``What is the adequate way of
averaging of the instability increment (if it exists)?" To answer this
question  one has to describe \emph{parametric instability  in spatially
inhomogeneous media}.

This problem has been discussed in various physical situations and
is presented in many books, for example in Sec. 6.5.2 of
monograph~\cite{NSW}, where explicit expressions for the
instability increment for different types of spatial
inhomogeneities are derived. Without going into details we can say
that the main physical message of this study is that for weakly
decaying parametric waves (which is our case) the threshold of the
parametric instability can be estimated with a good accuracy from
the total energy balance in the sample (cell, in our case).
Namely, at the threshold the total energy influx into the system
\begin{equation}
   \label{Wp}
W_+\propto \int  V[\bm r, \bm k(\bm r) ]\, d\bm r \,,
\end{equation}
   has to be equal to the total rate of the energy dissipation
$W_-$.  With the same prefactor as in \REF{Wp}
\begin{equation}
   \label{Wm}
W_-\propto \int   \gamma[\bm k(\bm r)]\, d\bm r \,, 
\end{equation}
   where $\gamma (\bm r)$ is the damping rate of the parametric waves.

The physical reason is that under wave propagation in weakly
inhomogeneous media its frequency serves as the adiabatic
invariant.  This  means that the wave frequency is independent of
the position, while the wave-vector $\bm k(\bm r)$ changes
accordingly to the dispersion law $\omega(\bm k, \bm r)=\,$const.
Because of that, even propagating in inhomogeneous media, the
waves with
\begin{equation}
\label{res}
\omega[\bm k(\bm r), \bm r]=\omega_L/2\,,
\end{equation}
   stay in parametric resonance with the pumping (homogeneous in
frequency, but spatially inhomogeneous spin  precession).  In
other words, if the mean-free pass of the wave exceeds the cell
size, the entire cell can be considered as the resonator with some
parametrical  mode, which is locally close to the planar spin
wave. Under this conditions, the threshold can  be well estimated
from the condition:
\begin{equation}
   \label{cond}
\int  V[\bm r, \bm k(\bm r) ]\, d\bm r= \int  \gamma[\bm k(\bm
r)]\, d\bm r \ .
\end{equation}

   Accuracy  of this estimate is about  $\pm (10  \div 30)\%$,
   and related with
ignoring in  \REF{cond}   the spatial variation of the actual
profile of the parametric waves (for more details, see i.e.
   Ref.~\cite{NSW}).

In experimental conditions, (see Fig.~1 in our Ref.~\cite{BLV})
one roughly says, that the fraction of the cell volume (near the
wall), with essential deviation from the planar geometry is about
of a half of the total volume. This is the volume, where the
energy pumping is dominating [i.e. the integral in the LHS of
\REF{cond}]. In this region, in agreement with the the third
comment of~\cite{Fcr}, the spin-wave vectors are relatively small
and thus one can neglect the wave damping. The wave damping [the
integral in the RHS of \REF{cond}] is dominated by the central
part of the cell, where wave vectors are sufficiently large.
Accordingly, the total energy balance in the cell is kept by the
spatial energy flux from the near-wall region to the central part
of the cell. Under this conditions, the threshold \REF{cond} can
be roughly rewritten in the simple manner:
\begin{equation}
   \label{cond1}
{\cal V}_V \max  V[\bm r, \bm k(\bm r) ] \simeq {\cal V}_\gamma
\max\gamma[\bm k(\bm r)]\,,
\end{equation}
   in spite of the fact, that the energy pumping and the energy
   damping are dominating in the different regions of the cell.
Here ${\cal V}_\gamma$ and ${\cal V}_V$ are the corresponding
effective volumes which depend on texture. As follows from the
analytical and numerical simulations, these volumes do not depend
on the value of magnetic field (see below) and in the considered
cell are roughly equal. This means that
   \begin{equation}
   \label{cond2}
    \max  V[\bm r, \bm k(\bm r) ] \simeq
\max\gamma[\bm k(\bm r)]\,,
\end{equation}
estimate, used in our Ref.~\cite{BLV}.

\noindent {\bf 3. Wave damping in the near-wall region}. In the
third complain  of~\cite{Fcr} we have been instructed  how to
estimate from \REF{res} wave vector of $\bm k(\bm r)$.  Presented
estimate shows that   near the wall wave vectors and,
consequently, wave damping is small with respect of the central
region with the conclusion that we miss a factor about 20 in our
estimate of the parametric threshold. The misunderstanding of
Ref.~\cite{Fcr} in this point is related to the question:
\emph{``How the value of the wave damping in the near-wall region
effects on the threshold of the parametric excitation in the
cell?"}  We hope that  our explanation in previous Sec.~2 is clear
enough  to realize  that the energy balance in the system of
weakly decaying waves has to be discussed rather globally, for the
entire cell, then locally, point-wise, as presented in
Comment~\cite{Fcr}. Therefore, the estimates of the spin-wave
vectors in the near-wall region, made in Com.~\cite{Fcr}, being
reasonable itself, are irrelevant to the problem at hand. The same
we can say about  the statement, made in~\cite{Fcr}, concerning
``additional factor $\approx 20$ in the RHS of Eq.~(28)" in our
Ref.~\cite{BLV}. As we explained in Sec.~2, the possible
inaccuracy of our estimate of the threshold~\cite{BLV} does not
exceed $(20\div30)\%$.

\noindent {\bf 4. Generalization  on  other precessing states}. To
conclude the subject of the spatial inhomogeneity we mention, that
the precession under discussion -- the HPD -- is a very specific
precessing state due to its unique symmetry. Only in the case of
the Brinkman-Smith mode the spin-orbit interaction does not
contribute to the amplitude of the parametric instability $V$, and
thus the parametric excitation by the BLV mechanism requires the
existence of the texture.

However, there are many other modes of spin precession in $^3$He,
for which the spin-orbit potential energy as a function of $S_z$
has also the concave form shape. These are: the so-called HPD2 in
$^3$He-B \cite{HPD2-B-phase}; the special mode of precession in
$^3$He-A \cite{HPD-A-phase}; the precession at one half of
equilibrium magnetization and at almost zero magnetization
observed in $^3$He-B \cite{HalfMagnetization}; etc. As distinct
from the HPD based on the Brinkman-Smith mode, in all these modes
the spin-orbit energy is oscillating, and thus it produces the
non-zero contribution to $V$  even in case of the spatially
homogeneous precession. This means that for all these modes of
spatially homogeneous precession,  the BLV mechanism, based on the
spin-orbit energy, will compete with the SF mechanism even at
moderate magnetic fields.

Notice, that some of the predicted HPD modes   still have not been
observed or identified. Now with our knowledge of the parametric
instability mechanisms and their dependence on different
parameters, we are able to find the region of parameters (magnetic
field, temperature, angles of precession, superfluid velocity,
etc.) where all
the conditions for the stability of precession are satisfied.\\

Now we are coming to the philosophical discussion, open in
Com.~\cite{Fcr} of ``validity of an idealization for a particular
experimental set-up" and  of  acceptance of ``generalization of
theory in a way, that makes possible its application to wider
class of experimental conditions".  Obviously, a success of
``idealization" or ``generalization", used in a theory, depends on
an experience,  physical intuition,  taste and courage  of its
authors, that are different for different investigators.  We hope
that we made it clear in   Secs. 1--3, that what it seems for the
author of Com.~\cite{Fcr}   ``ambiguous assumptions and non
justified approximations" have  indeed   clear sense and well
established physical background.

Nevertheless we are happy, using  the discussion with
Com.~\cite{Fcr}, to provide our theory of catastrophic
relaxation~\cite{BLV} with additional support from the  direct
numerical simulations of full Leggett-Takagi equations  \cite{BG}
(in the next Sec.~5) and especially from the laboratory study of
magnetic field dependence of the temperature of catastrophic
relaxation in $^3$He, Sec.~6. Note that most of the complications
related to the spatially inhomogeneous precession disappear at
high magnetic field, where the wave vector $k$ is almost constant
across the cell.

   \noindent
{\bf 5.~Discussion of numerical experiment}.   One-dimensional
numerical simulations of the Leggett-Takagi equations were made in
Ref.~\cite{BG}. It was found that in the viscous  limit, i.e. at
sufficiently high temperature,  the Brinkman-Smith configuration
takes place in the central part of the  cell being disturbed by the
periodic perturbations in angle $\theta$ coming from the
peripheral region, where the vector  $\bm L$ considerably deviates
from the Brinkman-Smith configuration.  This observation confirms
our statement, made in Sec.~1, that the spatial inhomogeneous
texture is stable, providing minimum of the spin-orbit
and  gradient energies at given boundary conditions and can
not be destroyed by any relaxation processes.

  At further cooling of the cell,  the BS configuration is more and more
disturbed, and at some  at some threshold temperature $T_{\rm th}$ the
catastrophe occurs: the exponential growth of  standing spin waves [with
frequency
$\omega_L/2$] starts in the peripheral region and propagates to the
central part, taking energy from the homogeneous precession and finally
destroying it. This picture is in full agreement with the
parametric instability of the inhomogeneous precession, studied in
Refs.~\cite{BLV,SF}.

Let us remind that there are two competing contributions to the
increment of the parametric instability in Eq.~(\ref{sum}). The
``intrinsic'' contribution $V_{_{\rm SF}}$ comes from the whole
sample; and the contribution $V_{_{\rm BLV}}$ which is due the
oscillating spin-orbit energy and it only comes from the regions
where the configuration deviates from the Brinkman-Smith mode. In
the one-dimensional geometry, used in the numerics~\cite{BG}, the
SF contribution  is suppressed in the central region because the increment
is zero for the waves radiated in the direction perpendicular  to  $\bm L$
(cf. Eq.(26) of Ref.\cite{SF} and Eq.(25) in Ref.~\cite{BLV}. This
means that while these particular numerical simulations support
the process discussed by the BLV theory~\cite{BLV}), they cannot
either support or disregard the relevance of   SF contribution
at given parameters of the simulations.

{\bf 6. Magnetic field dependence: SF \bf \emph{  vs.} BLV
competition}. To find out which contribution is dominating in real
experimental situations, we will  compare both contributions to
the increment~\Ref{sum}~\cite{BLV,SF} directly with laboratory
experiments. The BLV arguments~\cite{BLV} demonstrate that in the
moderate magnetic fields the  $V_{_{\rm BLV}}$ contribution  is
dominating. Both groups~\cite{Fcr,BLV} agree, that  according to
the calculations of Ref.~\cite{SF},  the SF contribution $V_{_{\rm
SF}}$
  dominates  at large magnetic field, when $\omega_L \gg
\Omega_L$. To see that this is really the case, we examine here
experimental data on magnetic field dependence, that include the
region of higher magnetic fields.

\begin{figure}
\includegraphics[height=7cm]{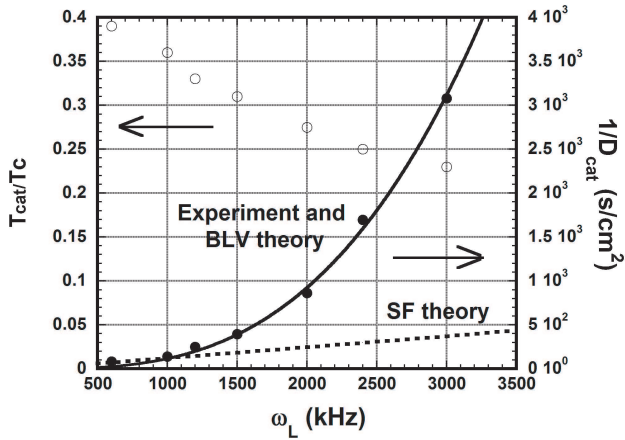}
\caption{ The experimental data for the temperature of catastrophic
relaxation as a function of the NMR frequency at 31 bar ( o ), left
scale, and the estimated value of the inverse spin diffusion
coefficient, (
    ) right scale. Solid line - BLV theoretical estimation~\cite{BLV},
dashed line - SF results~\cite{SF}.}

\end{figure}

The most relevant for this goal data  are   provided by the
   Cornell~\cite{Lee} and Grenoble~\cite{Bun} experiments,
where the temperature of catastrophic relaxation has been measured
as a function of magnetic field, see Fig.~1. Open circles in
Fig.~1 demonstrate the temperature $T_{\rm cat}(\omega_L)$ of the
catastrophic relaxation as a function of the magnetic field (or
Larmor frequency $\omega_L$) at 31 bar \cite{Lee}. The main effect
of the temperature is to provide the dissipation which damps the
parametric instability via the spin-diffusion mechanism \cite{SF}.
That is why the relevant physical quantity is the spin-diffusion
coefficient $D(T,\omega_L)$, rather than the temperature itself,
and we must convert the temperature to diffusion.   According to
Ref.  \cite{Einzel}, in the ballistic regime considered here when
$\omega_L\, \tau(T) \gg 1$,  the spin diffusion $D(T,\omega_L)
\sim 1/\omega_L \,\tau(T)$, where  $\tau(T)$ is a quasiparticles
scattering time, which grows exponentially with cooling. The
experimental temperature dependence of the spin diffusion is
obtained from measurements of diffusion through the HPD boundary
\cite{Dif}.

Combining these results with $T_{\rm cat}(\omega_L)$ from Fig.~1
one obtains the spin diffusion at which the
catastrophic relaxation occurs as a function of $\omega_L$: solid
circles in Fig.~1 show $D_{\rm cat}^{-1}(\omega_L)$.
    The experimental points fit excellently the cubic
dependence
\begin{equation}
  \frac1 { D_{\rm cat}^{^{\rm BLV }}}  =    \frac{\omega_L^3}
  { 4\bar a \, \Omega_L^2
c_\parallel^2}\propto \omega_L^3\,,
      \label{Cubic}
     \end{equation}
obtained within the BLV mechanism \{see  Eqs.~(23) and (24) in
Ref.~\cite{BLV}\}.

For the quantitative comparison of the experiment for   $
\omega_L$ dependence of  $1\big / D_{\rm cat}$   with Eq.
(\ref{Cubic}) we shall take the experimental values of the
parameters $\Omega_L$, $c_\parallel$ and use the parameter $\bar
a$, which characterizes the ${\bm L}$-texture, as a fitting
parameter. In the considered region of temperature all these
parameters are slow functions of temperature. We shall use the
data at 1 MHz and $T=0.35 T_{\rm c}$, where reliable experimental
data exist.
    From Ref.~\cite{Hakonen} we can estimate $\Omega_L^2 =
10^{11} Hz^2$; while from measurements of different modes of HPD
oscillations \cite{BDM} with pressure scaling by Fermi velocity,
we find $c_\parallel^2 = 1.5$ x $10^6$ cm$^2/$s$^2$. By
introducing these values into  Eq.~(\ref{Cubic}), one obtains
$\bar a = 0.07$, which is in good agreement with the theoretical
estimation of $\bar a \approx 0.1$ for 6 bar, made in
Ref.~\cite{BLV}.

In the current  consideration it is important that the parameter
$\bar a$ does not depend on magnetic field. The reason for that is
that according to the Ref. \cite{Wetting}, the characteristic
length scale of the near wall region is $\sim
c/\sqrt{\omega(\omega- \omega_L)}$. It is typically about
$c/\Omega_L$, and thus does not depend on $\omega_L$.

Next step is to compare the experimental results with the  SF
contribution \cite{SF}. It reads:
\begin{equation}
     \frac1 { D_{\rm cat}^{^{\rm SF }}} =
       \frac{\omega_L^2}
     { 2 \lambda_{\rm max}\,
c_\parallel^2}\propto \omega_L\ .
      \label{Demp2}
     \end{equation}
Here we accounted for that  $\lambda_{\rm max}$ near the HPD
boundary is equal $0.016 \omega_L$ and therefore  the field
dependence of $1/ D_{\rm cat}^{^{\rm SF }}$ should be linear. As
one see in Fig.~1 this    clearly  contradicts to the
experiment. Furthermore, if we plot the value of $1/ D_{\rm
cat}^{^{\rm SF }} $, we find (Fig.~1) that the SF result agrees
with   the experiment only in the region of NMR of 500 - 1000 MHz.
At higher fields the theoretical value of diffusion, at which
catastrophic relaxation with the SF mechanism  should occur,
definitely disagrees with  the experiments.

   \noindent

{\bf Discussion}. According to our theoretical analysis, in
moderate magnetic fields, i.e. at $\omega_L$ smaller than about 1
MHz, the BLV contribution the parametric instability of HPD is
dominating.  At these fields the effect of spatially inhomogeneity
on the BLV mechanism is essential, and  we clarified in Secs.~2
and 3 the corresponding points, rouse   in Com.~\cite{Fcr}. Most
of the complications related to the issue of spatial inhomogeneity
do not arise at high magnetic fields, where $\omega_L\gg
\Omega_L$, and the wave vector $k$ is (almost) homogeneous along
the cell. However, we expected that at such a high field the BLV
contribution is subleading , while  the SF contribution dominates,
if $\omega_L^2>10~ \Omega_L^2$. The same opinion was expressed in
   Com.~\cite{Fcr}, where it was stressed that the stronger
magnetic fields have to be used for the experimental investigation
of the ``intrinsic'' SF mechanism  of catastrophic relaxation.

On the contrary, the surprising  experimental fact is that the
magnetic field  dependence of the catastrophic relaxation
demonstrates that even up to a rather high field, when
$\omega_L^2\sim 100~ \Omega_L^2$,  it is still quantitatively and
qualitatively described by the BLV contribution to the parametric
instability. Moreover,  the magnetic field dependence is in
striking disagreement with the SF contribution. The matter of the
fact is that the SF contribution does not show up in these
experiments at all.

We do not think that explanation of this fact is related to
possible calculational mistakes in the analytics, presented in
Ref. \cite{SF}. We feel that   reason(s) for the obvious
qualitative disagreement between the SF analysis and experiment is
deeper and may be related to possible violations of the SF
approach in the region of large $k$ vectors.

To make long story short, the BLV mechanism of the parametric
instability   gives quantitative description of the present experiments
for any values of $\omega_L$ used, while the SF contribution, that seems
to be the leading one (for large $\omega_L $), is absent in the
experiments by unknown reason. This is the main puzzle; its
solution requires the further theoretical, numerical and
experimental efforts, that are beyond the scope of this Letter.

   \noindent

{\bf Acknowledgements}. We acknowledge the  critical comments by
I.A.~Fomin~\cite{Fcr}, that inspire us for further  clarification
and development of our theory of catastrophic relaxation in
superfluid $^3$He-B, presented in this Letter. This work was done
as the result of collaboration in the framework of the ESF Program
COSLAB, the Large Scale Installation Program ULTI of the European
Union (contract number: RITA-CT-2003-505313) and the project
ULTIMA of the "Agense National de Recherche", France.  The work
was also supported in part by the Russian Foundation for
Fundamental Research and the US-Israel Binational Science
Foundation.

\end{document}